\documentclass[reprint,showpacs,amsmath,amssymb,aps,prb,floatfix,lengthcheck]{revtex4-1}

\usepackage{graphicx}% Include figure files
\newcommand{\bvec}{\mathbf}

\begin{document}

\title{Off-resonant all-optical switching dynamics in a ferromagnetic model system}

\author{Christiane Scholl}
\affiliation{Physics Department and Research Center OPTIMAS, Kaiserslautern University,
P. O. Box 3049, 67663 Kaiserslautern, Germany}
\author{Svenja Vollmar}
\thanks{Graduate School of Excellence Materials Science in Mainz, Gottlieb-Daimler-Strasse
47, 67663 Kaiserslautern, Germany}
\affiliation{Physics Department and Research Center OPTIMAS, Kaiserslautern University,
P. O. Box 3049, 67663 Kaiserslautern, Germany}
\author{Hans Christian Schneider}
\email{hcsch@physik.uni-kl.de}
\affiliation{Physics Department and Research Center OPTIMAS, Kaiserslautern University,
P. O. Box 3049, 67663 Kaiserslautern, Germany}

\date{\today}
\begin{abstract}
We present a theoretical study of the the effects of off-resonant polarized optical fields on a ferromagnetic model system. We determine the light-induced dynamics of itinerant carriers in a system that includes magnetism at the mean-field level and spin-orbit coupling. We investigate an all-optical switching process for ferromagnets, which is close to the one proposed by Qaiumzadeh et al. [Phys. Rev. B 88, 064416] for the inverse Faraday effect. By computing the optically driven coherent dynamics together with incoherent scattering mechanisms we go beyond a perturbation expansion in powers of the optical field. We find an important contribution of a dynamic Stark effect coupling of the Raman type between the magnetic bands, which leads to a polarization-dependent effect on the magnetization that may support or oppose switching, but also contributes to demagnetization via an increase in electronic energy. 
\end{abstract}


\maketitle

\emph{Introduction.} The optical excitation of magnetic systems has many facets that have been explored over the last two decades. After demagnetization of 3d-ferromagnets by ultrashort pulses was discovered,~\cite{Beaurepaire.1996} it was realized that in alloys with anti-ferromagnetically coupled sublattices the excitation by ultrashort pulses can lead to a transient ferromagnetic-like state~\cite{Radu.2011} and even a complete reversal of the magnetization. These magnetization dynamics can be understood in terms of transient heating effects.~\cite{Mentink.2012,Ostler.2012,Wienholdt.2013,Baral:2015fu} More recently, there has also been evidence for magnetization switching induced in \emph{ferromagnets}~\cite{Mangin2014-bx,John2017-mk} where a purely heat-induced effect should not work. In this case, there should be a microscopic mechanism with which the polarized optical fields act on the magnetization. Currently most popular explanation is the inverse Faraday effect (IFE), which has been studied in the last decade for ultrafast magnetism.~\cite{Kimel2005-lq,Kirilyuk2010-wx,Alebrand2012-dj,Hassdenteufel2015-om} However, the name IFE is applied to different mechanisms and models of light-matter interaction, some of which are more or less classical in nature.~\cite{Pitaevskii1961-qd,Pershan1966-wf,Hertel.2006,Kurkin2008-js,Woodford2009-kf,Taguchi2011-bi,Hertel2015-oa,Edelstein1998-lw,Qaiumzadeh2016-pd} The IFE was introduced for paramagnetic materials, but there are important differences for this effect between paramagnetic and ferromagnetic materials.~\cite{Zheludev1994-im,Mikhaylovskiy2012-vs} 

Recent quantum mechanical calculations of the time-dependent IFE for ferromagnets have been performed in different ways. A perturbation theory in orders of the external field for band ferromagnets suggests that the magnetization is most efficiently influenced directly by a coherent optical field if this field is resonant with dipole transitions in the system.~\cite{Battiato2014-mu,Berritta2016-lz} Calculations for few-level systems have analyzed a stimulated Raman effect, where the field is resonant with intermediate levels~\cite{Popova2012-zl} and where angular momentum is transferred from the field. Qaiumzadeh et al.~\cite{Qaiumzadeh2013-cy} proposed the spin-dependent optical Stark effect as a mechanism for the inverse Faraday effect that works for \emph{off-resonant} optical fields. In this scenario the spin-dependent optical Stark effect changes the splitting between the spin-up and spin-down bands  and the actual magnetization change occurs by redistribution of electrons between the split bands so that the lattice acts as source and sink of angular momentum. 

The present paper analyzes a mechanism for all-optical switching by off-resonant fields that is based on the ideas of Qaiumzadeh et al.~\cite{Qaiumzadeh2013-cy} but modifies and extends their model to also include angular momentum exchange with the optical field and spin-orbit coupling in the magnetic bands. It combines the action of a coherent optical field, incoherent scattering dynamics and $k$-dependent spin-orbit contributions in a ferromagnetic model band structure of itinerant electrons. The main purpose of this paper is to elucidate general properties of the magnetic switching dynamics due to the interaction with off-resonant optical fields. In the following, we will generally refer to the optically induced magnetization dynamics as all-optical switching and avoid the term ``inverse Faraday effect,'' because of the different meanings attached to it.~\cite{Pitaevskii1961-qd,Pershan1966-wf,Hertel.2006,Kurkin2008-js,Woodford2009-kf,Taguchi2011-bi,Popova2012-zl,Hertel2015-oa,Battiato2014-mu,Berritta2016-lz}

\begin{figure}[t]
\centering
\includegraphics[width=0.4\textwidth]{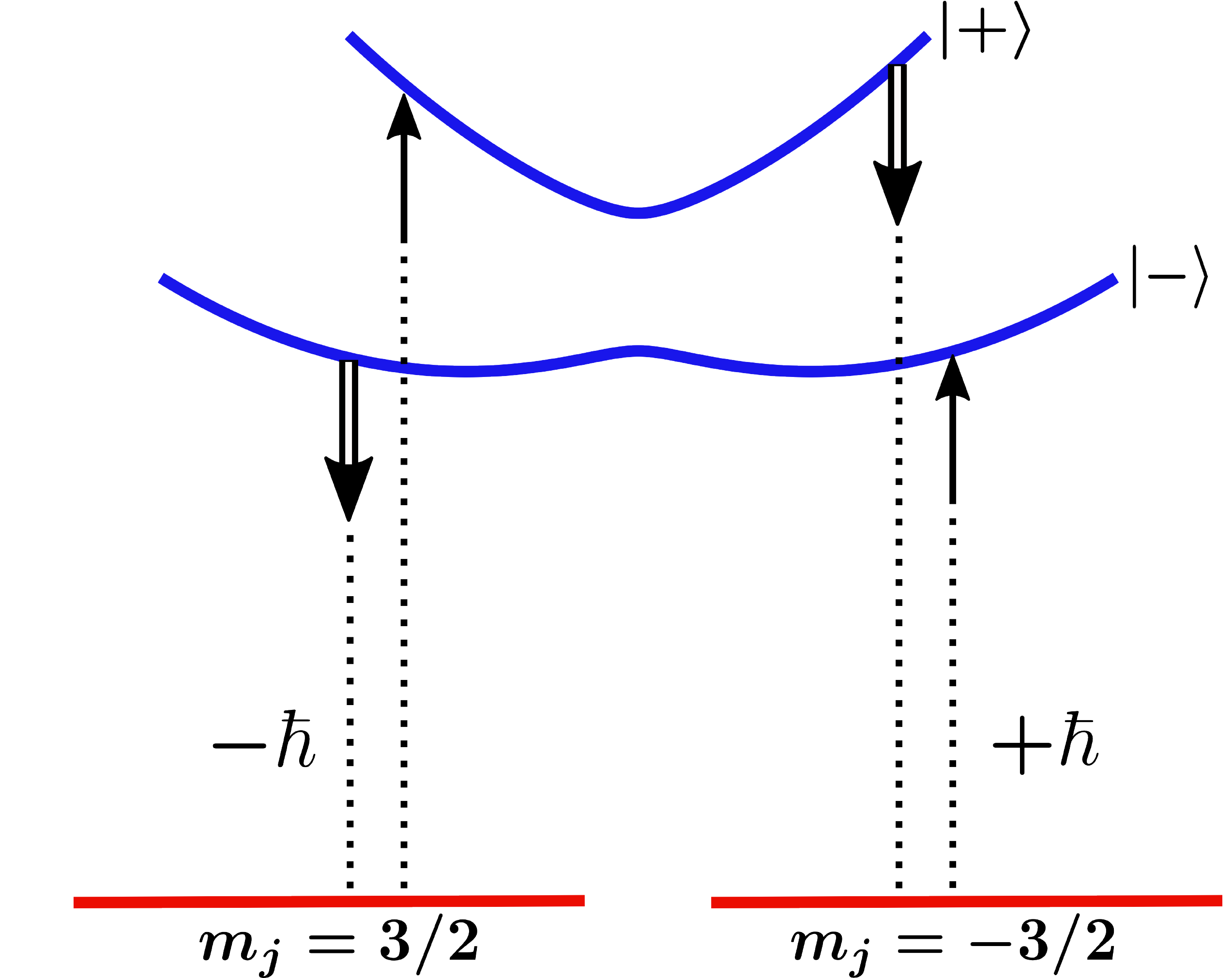}
\caption{Schematic band structure and transitions. The bands $|+\rangle$ and $|-\rangle$ are s-like bands with a magnetic splitting due to a Stoner mean field and are coupled by electric dipole transitions (dotted lines) to two p-like bands. The different dipole coupling matrix elements are indicated by the thickness of the arrows and the angular momentum transferred to the carriers by a left (right) circularly polarized photon is indicated by $-\hbar$($+\hbar$). Both $|+\rangle$ and $|-\rangle$ states are connected to each of the p-like state by each circular optical field polarization. The arrows for the different transitions are offset to show the Raman-like nature of the scheme, there is no momentum transfer.}
\label{fig:bands}
\end{figure}

\emph{Model.} We intend to capture important aspects of the mechanisms underlying all-optical switching by taking into consideration as essential ingredients spin-orbit-coupling (SOC), electronic scattering processes and a ferromagnetic exchange splitting that can change in response to the electronic dynamics and thus captures time-dependent magnetic properties. The model band structure is shown in Fig.~\ref{fig:bands}. We consider two partially filled s-like bands $|\pm,\bvec{k}\rangle$, which exhibit a magnetic splitting of about 20\,meV due to a Stoner mean-field contribution but also include spin-orbit coupling. For the purpose of nonlinear optics the magnetic bands are the ``essential bands'' and are dipole coupled to two ``non-essential bands'' $|\pm\frac{3}{2},\bvec{k}\rangle$, which are modeled as filled p-like bands (with total angular momentum $j=1$ and $m_j=\pm3/2$ and effective mass~$0.4m_{\text{e}}$)~\cite{haug2009quantum} and are separated from the essential magnetic bands by an energy that is large compared to the equilibrium magnetic splitting. In order to keep the numerical calculations manageable, we use a Rashba-type spin-orbit coupling with an effectively 2-dimensional $k$~space and include magnetism at the level of a Stoner splitting
\begin{equation}
\hat{H} = \begin{pmatrix}
\frac{\hbar^{2} \bvec{k}^{2}}{2 m^{*}} + \frac{U}{2} (n-m)
& - \alpha (k_y + i k_x) \\
\alpha (- k_y + i k_x) 
& \frac{\hbar^{2} \bvec{k}^{2}}{2 m^{*}} + \frac{U}{2} (n+m)
\end{pmatrix} \ 
\label{eq:hamiltonian}
\end{equation}
where $n$ denotes the particle density and $m$ represents the spin polarization per site, which is determined from the reduced density matrix, see below. The Stoner exchange splitting is $\Delta = U m$, and the magnetic and spin-orbit splitting are controlled, respectively, by the Stoner and Rashba parameters $U$ and $\alpha$. For our model system, the carrier densities and splitting are smaller than in metallic ferromagnets, but they are consistent with each other, and the model includes the complete carrier and magnetic dynamics of the switching process. In particular, if the carrier distributions in the electronic bands change due to carrier scattering and/or interaction with optical fields, this affects the magnetization and electron density, so that the instantaneous energy dispersions described by~\eqref{eq:hamiltonian} also change with time. 

The magnetization dynamics are induced by the coupling to the optical field via dipole matrix elements connecting the magnetic bands and the non-essential bands, as shown in Fig.~\ref{fig:bands}. In difference to Qaiumzadeh et al.,~\cite{Qaiumzadeh2013-cy} the Rashba spin mixing in the magnetic states means that each circular polarization state of the optical field couples both magnetic bands to a p-like state, which makes Raman-like transitions between the magnetic states possible, as indicated by arrows in Fig.~\ref{fig:bands}. While the directions of the transitions depend on the system parameters and excitation conditions, there is an asymmetry between opposite circular field polarizations due to the different strengths of the dipole matrix elements between the magnetic states and p-like states, as signified by the thickness of the arrows in Fig.~\ref{fig:bands}. We describe the coherent dynamics of the 4 band system sketched in Fig.~\ref{fig:bands} using the reduced single-particle density matrix $\rho_{\bvec{k}}^{\mu \nu} = \langle \hat{c}_{\bvec{k},\nu}^{\dagger} \hat{c}^{\phantom{\dagger}}_{\bvec{k},\mu} \rangle$, where $\hat{c}^{\dagger}_{\mu,\bvec{k}}$ creates a particle in a Bloch state labeled by band index $\mu$, which runs over the 4 bands shown in Fig.~\ref{fig:bands}, and crystal momentum $\bvec{k}$ :
\begin{equation}
\begin{split}
 \frac{\partial}{\partial t}& \rho_{\bvec{k}}^{\nu \mu}
  = \frac{i}{\hbar} \left(\epsilon_{\mu} (\bvec{k}) - \epsilon_{\nu} (\bvec{k}) \right) \rho_{\bvec{k}}^{\mu \nu} \\
  &+ \sum_{\mu'} \big[ \rho_{\bvec{k}}^{\nu\mu'} \Omega_{\bvec{k}}^{\mu', \mu} (t)  - \Omega_{\bvec{k}}^{\nu, \mu'} (t)\rho_{\bvec{k}}^{\mu' \mu}  \big]  +  \frac{\partial}{\partial t} \rho_{\bvec{k}}^{\mu \nu}\Big|_{\text{scat}} ,
\end{split}
\label{eq:OBEs}
\end{equation}
Here, $\Omega_{\bvec{k}}^{\mu\nu} (t)= \bvec{d}_{\mu,\nu}(\bvec{k})\cdot \bvec{E}(t)$ are the matrix elements of the Rabi energy, which contains the real electric field vector $\bvec{E}(t)$ and the matrix elements of the dipole operator $\bvec{d} = - e \bvec{r}$.  

We assume that these vanish between the magnetic bands and use the Rabi energy as a measure of the field amplitude. Importantly, the band energies $\epsilon_{\mu}(\bvec{k})$ and basis states $|\mu, \bvec{k}\rangle$ used for the matrix elements in~\eqref{eq:OBEs} are the instantaneous eigenenergies of the mean-field hamiltonian~\eqref{eq:hamiltonian}. In each timestep, the basis of single particle states $|\mu,\bvec{k}\rangle$ corresponding to the instantaneous values of $n$ and $m$ is used. This procedure includes the influence of the coherent optical field and the change of the instantaneous quasiparticle band structure via incoherent redistribution and relaxation processes.

We compute the incoherent redistribution of carriers, i.e., scattering, only for the magnetic bands, as we assume that only this is relevant for the magnetization dynamics. To this end we introduce scattering contributions to the elements $\rho^{\mu\nu}$ with $\mu,\nu=\pm$ of Eq.~\eqref{eq:OBEs}, in the form of a relaxation-time ansatz designed for the treatment of systems with spin-orbit coupling and mean-field magnetism. We use a generalized relaxation time ansatz  $d\rho_{\bvec{k}}^{\mu\nu}/dt|_{\text{scat}}=-\frac{\rho_{\bvec k}^{\mu \nu}-\tilde{\rho}_{\bvec k}^{\mu \nu}}{\tau}$ in~\eqref{eq:OBEs}, where $\tilde{\rho}_{\bvec k}^{\mu \nu}$ indicates the elements of a suitably determined quasi-equilibrium spin density matrix.~\cite{Scholl-unpublished} In order to model electron-electron scattering, we determine $\tilde{\rho}$ such that the relaxation-time ansatz conserves spin polarization, energy density and charge density of the carriers in the essential bands. The quantity $\tau$ plays the role of an effective scattering time for the incoherent electron-electron-scattering. The spin polarization of the magnetic bands, which we take as the magnetization, is calculated via $ \bvec{m} =  \sum_{\bvec{k}, \nu, \nu'} \langle \nu , \bvec{k} | \vec \sigma | \nu' , \bvec{k} \rangle \rho^{ \nu\nu'}_{\bvec{k}} $ and includes the time-dependent spin expectation values of the  eigenstates~$|\nu,\bvec{k}\rangle$ of~\eqref{eq:hamiltonian}.

\begin{figure}[t]
\centering
\includegraphics[width=0.45\textwidth]{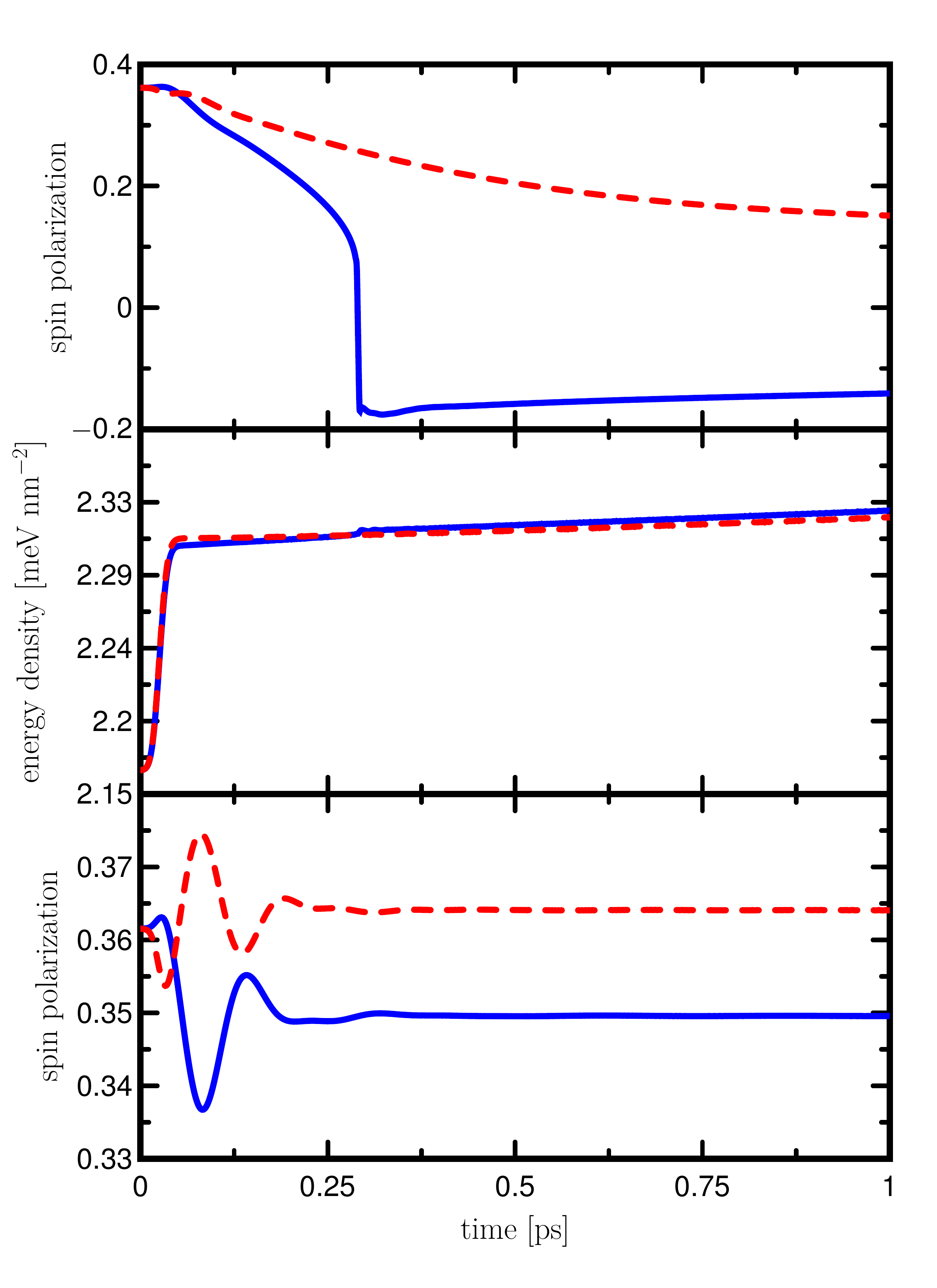}
\caption{(a) Time dependent magnetization (spin polarization) of the ferromagnetic model system excited by a optical field with right-circular (dashed line) and left-circular (solid line). Magnetization switching occurs only for left-circularly-polarized excitation. (b) corresponding energy density.  (c) Same calculation as (a) without electronic redistribution.}
\label{fig:offresonant}
\end{figure}

\emph{Results.} We determine a self-consistent magnetic ground-state for the model~\eqref{eq:hamiltonian} with initial magnetization direction $+z$. For the calculations presented in the following, we assume a Stoner parameter of 50 meV, a Rashba parameter of $20$\,meV\,nm, $m^{*}=m_{\text{e}}$, and a equilibrium temperature for the mean-field ground state calculation of $T=70$\,K. Starting from this ground state, we compute the dynamical distribution functions and states by solving~\eqref{eq:OBEs} together with ~\eqref{eq:hamiltonian}. The excitation is characterized by coupling to the non-essential bands, which are separated by $1.5$\,eV. For the effective electron-electron scattering time we take $\tau = 50$\,fs.

We consider the switch-on of a CW field which is ramped up with a rise time of $40$\,fs at $t=0$ and to an amplitude corresponding to a Rabi energy of $\hbar\Omega=15$\,meV, for a transition between a non-essential band state and a magnetic band as shown in Fig.~\ref{fig:bands}. The frequency of the optical field is detuned by $200$\,meV with respect to the transition to bottom of the lower band. We always assume a vanishing dephasing contribution, i.e, $d\rho^{\mu\nu}/dt|_{\text{scat}}\equiv \gamma\to 0$ for the off-diagonal elements of the density matrix corresponding to the optical transitions, which is well justified for a large detuning.~\citep{haug2009quantum} Fig.~\ref{fig:offresonant}(a) shows the time-dependent magnetization including scattering contributions for off-resonant left polarized and right polarized optical fields. For the left-polarized optical field the magnetization is first reduced to about 1/4 of its initial value, and then the magnetization direction is reversed rapidly in about 50\,fs. After magnetization switching is completed at about 250\,fs, there is a slower decay of the magnetization. For right polarized light we find in Fig.~\ref{fig:offresonant}(a) essentially only demagnetization. Comparison of the two light polarization states shows that this setup realizes off-resonant all-optical switching in a model ferromagnet. 

To explain the behavior of the magnetization for the different polarization states of the optical field, we take a closer look at how the optical fields influence the electronic dynamics in the spin-orbit coupled essential bands. In addition to the magnetization, we characterize the electronic dynamics by the density and total energy density $e=\sum_{\bvec{k}\mu}\epsilon_{\bvec{k}\mu}\rho^{\mu\mu}_{\bvec{k}}+U n_{\uparrow}n_{\downarrow}$. The latter quantity is shown in Fig.~\ref{fig:offresonant}(b). When the optical field is switched on, the energy increases rapidly and subsequently shows a slow increase. The comparatively large initial increase by 5\% is due to our choice of model system whose density of states is smaller than in a metallic ferromagnet. The increase in energy is accompanied by an 0.5\% increase of the electronic density in the magnetic bands (not shown).

The initial change in the energy and electronic density is due to the 40\,fs switch-on of the optical fields, which leads to Fourier components of the field that are resonant with transitions between the filled bands and the magnetic bands, even though we use a vanishing dephasing $\gamma \to 0$ of the optical polarization. One could call this a stimulated Raman process. We stress that this stimulated Raman process involves photons of the same polarization, so that it is fundamentally different from the process studied by Popova et al.~\cite{Popova2011-jr}

In addition to the initial rise of electronic energy in the magnetic bands with the switching-on of the field, there is a subsequent slower increase in electronic energy visible in Fig.~\ref{fig:offresonant}(b) for both right and left-polarized optical fields. After the switch-on, the field is essentially continuous-wave, so that there is no stimulated Raman transition and the influence of the optical field is better described as dynamic Stark effect coupling of the Raman type.~\cite{Sussman.2011} These Raman-like transitions are indicated by arrows in Fig.~\ref{fig:bands}. This is similar to the mechanism suggested by Qaiumzadeh et al., but in their model the electronic bands are not spin mixed and the Raman transition is between the non-essential bands instead of the magnetic electron bands in our case.

To assess the influence of the incoherent contributions to the dynamics, we compare Fig.~\ref{fig:offresonant}(a) with Fig.~\ref{fig:offresonant}(c), which shows the results if we repeat the calculation of Fig.~\ref{fig:offresonant}(a) with identical parameters but without scattering/carrier redistribution contributions. In this case there is a small change of spin polarization, i.e., the magnetization, in opposite directions for the different light polarizations (the difference will be discussed below), which proves that the direct effect on the magnetization induced by the polarized coherent optical fields is indeed small and incoherent scattering is needed for all-optical magnetization switching to occur. Without scattering, the charge and energy dynamics, which are not shown here, are very similar from the case with scattering shown in Fig.~\ref{fig:offresonant}(b). The carrier redistribution thus converts the energy increase due to the dynamic Stark effect coupling into a demagnetization, as suggested by Qaiumzadeh et al.~\cite{Qaiumzadeh2013-cy} In our microscopic calculation it does this via the interplay of $k$-dependent spin-orbit coupling and a spin conserving scattering processes.~\cite{Krauss.2009,Mueller.2013,Leckron.2017} Together, they act as an Elliott-Yafet-type process, which leads to a demagnetization whenever the energy in the electronic system is increased. Only when this demagnetization process is included in the calculation we find a pronounced difference between the magnetization dynamics for the different polarization states of the optical field. 

We calculate the whole magnetization dynamics, which leads to a reversal of the magnetization direction for the left-polarized optical field and which is due to a combination of the coherent field, the change in the band structure, and the incoherent redistribution dynamics.  Thus one cannot simply determine an effective magnetic field that is responsible for the switching dynamics, but we can compare to the calculation without carrier redistribution (scattering) processes, which is shown in Fig.~\ref{fig:offresonant}(c). In this case, the left-polarized light leads to a steady state with reduced magnetization, whereas the right-polarized optical field leads to a steady state with slightly \emph{increased} magnetization. We therefore find that for the left polarized optical field, the dynamic Stark-effect coupling works in the direction of magnetization reversal, but is not sufficient to switch the magnetization by itself. Only when the magnetic splitting is reduced so that the band structure is closer to that of a pure Rashba system, the optical field can switch the direction of the magnetization by reversing the small $z$ component of the single-particle expectation value $\langle \mu \bvec{k}|\sigma_z|\mu\bvec{k}\rangle$ of the  $\mu=+$ and $\mu=-$ bands, respectively.~\cite{Scholl-unpublished} For the right-polarized optical field the dynamic Stark-effect coupling alone would give rise to an increase in magnetization, but the concomitant increase in energy effectively leads to demagnetization. While we cannot compare this result directly to the measurement of El Hadri et al.\cite{El_Hadri2016-xa}, it shows qualitatively they correct behavior as the experiment also finds an effective demagnetization for the light polarization opposite to the one that leads to switching. The last observation in connection with Fig.~\ref{fig:offresonant} concerns the behavior for longer times ($>1$\,ps). There the  magnetization dynamics for the left and right polarized pulses become symmetric because for the right polarized field, the helicity of the light and the magnetization direction are aligned for the whole dynamics and for the left polarized case the helicity an magnetization are aligned in the same $-z$ direction. 

\begin{figure}[t]
\centering
\includegraphics[width=0.4\textwidth]{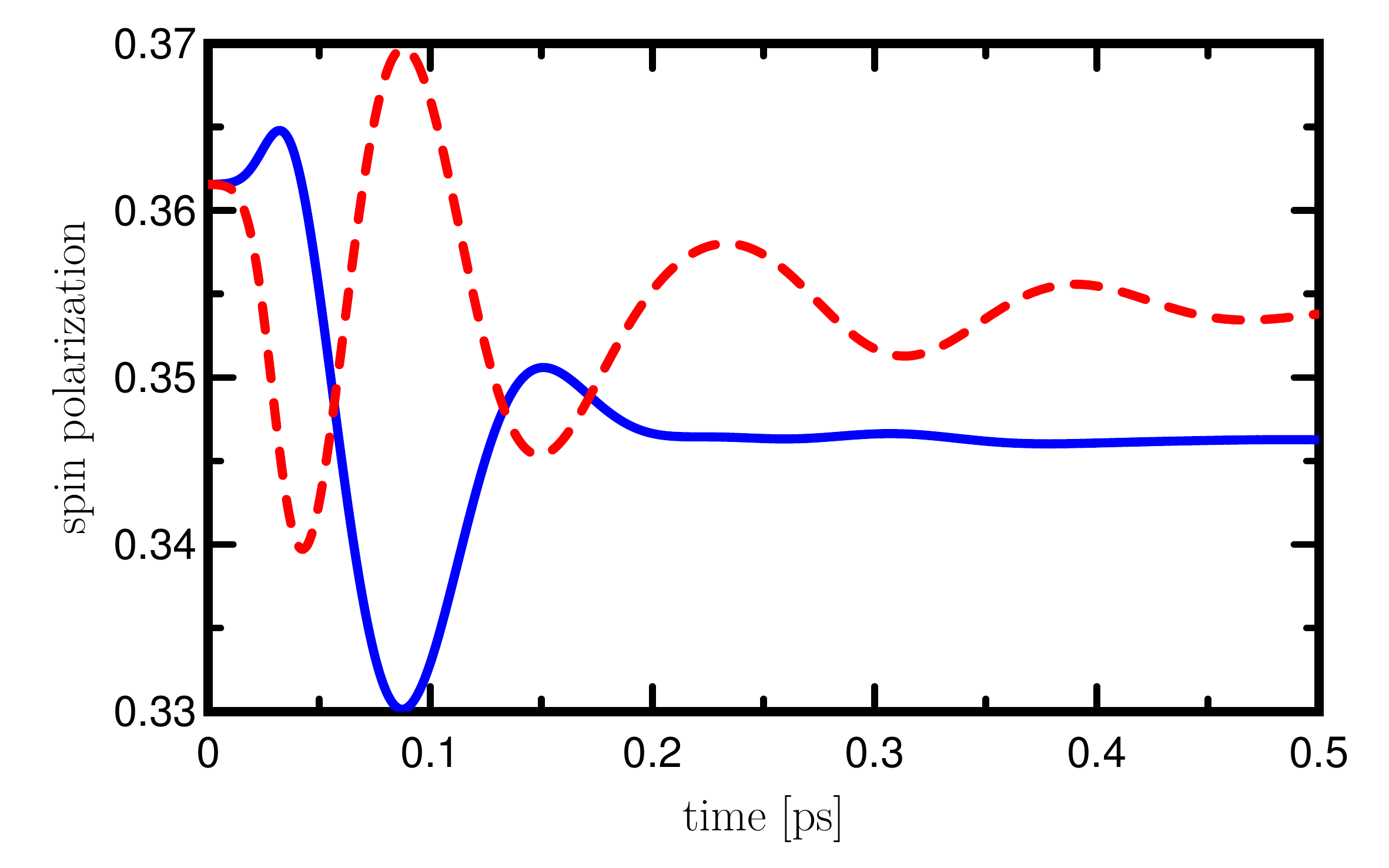}
\caption{Magnetization (spin polarization) dynamics computed without carrier redistribution and for a fixed band structure with a left-polarized (solid line) and a right-polarized (dashed line) optical field. The CW optical field is closer to resonance with a detuning of 5\,meV. This setup is designed to approximate the change of the density matrix due to the coherent optical field and leads to a reduction in magnetization for both circular polarizations of the optical field. }
\label{fig:ife_only}
\end{figure}

Finally, we would like to investigate the connection to perturbation theory with respect to the external field, which is often regarded as the sole contribution of the coherent optical field. We can achieve this by computing the magnetization dynamics for a fixed band structure. In this case, we do not update  the eigenenergies and eigenstates of hamiltonian~\eqref{eq:hamiltonian}, i.e., we keep the Stoner mean-field spliting fixed, so that the change in magnetization comes from the dynamics of the reduced density matrix under the influence of the optical field. Such a calculation for right and left-circularly polarized optical field is shown in Fig.~\ref{fig:ife_only}.  For a detuning of 200\,meV the effect on the magnetization is vanishingly small, so that, in addition to keeping the band structure fixed, we now use a field closer to the resonance between magnetic and non-essential bands, namely a detuning of 5\,meV. The other parameters are kept the same as in Fig.~\ref{fig:offresonant}. In this scenario, the magnetization approaches different steady states with a reduced magnetization (compared to equilibrium) for both right and left-polarized optical fields. As the density matrix includes the influence of the coherent field to all orders of the field, this result should correspond to a mechanism for the inverse Faraday effect analyzed by Oppeneer and coworkers in the framework of 2nd-order perturbation theory for close-to-resonance fields. Indeed, Berritta et al.~\cite{Berritta.2016} have also found that the influence of optical fields with opposite circular polarization may lead to a reduced steady-state magnetization with respect to the equilibrium magnetization. This is somewhat contrary to what one expects from older theories of the inverse Faraday effect which yielded exactly opposite effective magnetic fields for opposite circular polarizations of the optical field. This ``antisymmetry'' between the magnetic field and the circular optical polarization is not present because of the finite equilibrium magnetization, which breaks the symmetry between the dynamics induced by right and left-circularly polarized fields.~\cite{Agranat.1986} In our case, we find that the asymmetry is pronounced for the close-to-resonance case in Fig.~\ref{fig:ife_only} because there the optical field ``sees'' the ferromagnetic splitting. In the off-resonant case in Fig.~\ref{fig:offresonant}(c), the ferromagnetic splitting should play a smaller role, and indeed the magnetization dyamics are closer to realizing antisymmetry, as the magnetization is changed in opposite directions for opposite polarizations.

\emph{Conclusion.}  We introduced a microscopic dynamical model to study the all-optical magnetization switching process in a simple ferromagnetic band structure, including spin-orbit coupling and incoherent carrier redistribution/scattering processes. For off-resonant excitation we found that the switching process is a combination of demagnetization and the influence of the off-resonant field in the form of a dynamical Stark effect. The main angular momentum change is supplied by the lattice via an Elliott-Yafet like magnetization change, which results from the combination of spin-orbit coupling and scattering processes. Even for a continuous-wave excitation, we found that the field acts directly on the magnetization via an \emph{off-resonant} Raman-like processes, which is closely related to the dynamical Stark effect. This Raman-like process leads to a decrease/increase of the electronic spin polarization, i.e., the magnetization, as one would expect from the angular momentum supplied by the left/right circularly optical field. However, both circular polarizations increase the energy of the electrons in the spin-split bands during the duration of the optical field, and thus contribute to a demagnetization effect largely independent of the polarization. For optical fields close to resonance we find magnetization changes that are not antisymmetric with respect to the optical polarization, in agreement with recent perturbation theory calculations.

Svenja Vollmar received a fellowship through the Excellence Initiative (DFG/GSC 266). We acknowledge support from DFG by the SFB/TRR Spin+X.

\bibliographystyle{apsrev4-1}
%\bibliography{switching,SwitchingBib.bib}

\begin{thebibliography}{37}%
\makeatletter
\providecommand \@ifxundefined [1]{%
 \@ifx{#1\undefined}
}%
\providecommand \@ifnum [1]{%
 \ifnum #1\expandafter \@firstoftwo
 \else \expandafter \@secondoftwo
 \fi
}%
\providecommand \@ifx [1]{%
 \ifx #1\expandafter \@firstoftwo
 \else \expandafter \@secondoftwo
 \fi
}%
\providecommand \natexlab [1]{#1}%
\providecommand \enquote  [1]{``#1''}%
\providecommand \bibnamefont  [1]{#1}%
\providecommand \bibfnamefont [1]{#1}%
\providecommand \citenamefont [1]{#1}%
\providecommand \href@noop [0]{\@secondoftwo}%
\providecommand \href [0]{\begingroup \@sanitize@url \@href}%
\providecommand \@href[1]{\@@startlink{#1}\@@href}%
\providecommand \@@href[1]{\endgroup#1\@@endlink}%
\providecommand \@sanitize@url [0]{\catcode `\\12\catcode `\$12\catcode
  `\&12\catcode `\#12\catcode `\^12\catcode `\_12\catcode `\%12\relax}%
\providecommand \@@startlink[1]{}%
\providecommand \@@endlink[0]{}%
\providecommand \url  [0]{\begingroup\@sanitize@url \@url }%
\providecommand \@url [1]{\endgroup\@href {#1}{\urlprefix }}%
\providecommand \urlprefix  [0]{URL }%
\providecommand \Eprint [0]{\href }%
\providecommand \doibase [0]{http://dx.doi.org/}%
\providecommand \selectlanguage [0]{\@gobble}%
\providecommand \bibinfo  [0]{\@secondoftwo}%
\providecommand \bibfield  [0]{\@secondoftwo}%
\providecommand \translation [1]{[#1]}%
\providecommand \BibitemOpen [0]{}%
\providecommand \bibitemStop [0]{}%
\providecommand \bibitemNoStop [0]{.\EOS\space}%
\providecommand \EOS [0]{\spacefactor3000\relax}%
\providecommand \BibitemShut  [1]{\csname bibitem#1\endcsname}%
\let\auto@bib@innerbib\@empty
%</preamble>
\bibitem [{\citenamefont {Beaurepaire}\ \emph {et~al.}(1996)\citenamefont
  {Beaurepaire}, \citenamefont {Merle}, \citenamefont {Daunois},\ and\
  \citenamefont {Bigot}}]{Beaurepaire.1996}%
  \BibitemOpen
  \bibfield  {author} {\bibinfo {author} {\bibfnamefont {E.}~\bibnamefont
  {Beaurepaire}}, \bibinfo {author} {\bibfnamefont {J.-C.}\ \bibnamefont
  {Merle}}, \bibinfo {author} {\bibfnamefont {A.}~\bibnamefont {Daunois}}, \
  and\ \bibinfo {author} {\bibfnamefont {J.-Y.}\ \bibnamefont {Bigot}},\
  }\href@noop {} {\bibfield  {journal} {\bibinfo  {journal} {Physical Review
  Letters}\ }\textbf {\bibinfo {volume} {76}},\ \bibinfo {pages} {4250}
  (\bibinfo {year} {1996})}\BibitemShut {NoStop}%
\bibitem [{\citenamefont {Radu}\ \emph {et~al.}(2011)\citenamefont {Radu},
  \citenamefont {Vahaplar}, \citenamefont {Stamm}, \citenamefont {Kachel},
  \citenamefont {Pontius}, \citenamefont {D{\"u}rr}, \citenamefont {Ostler},
  \citenamefont {Barker}, \citenamefont {Evans}, \citenamefont {Chantrell}
  \emph {et~al.}}]{Radu.2011}%
  \BibitemOpen
  \bibfield  {author} {\bibinfo {author} {\bibfnamefont {I.}~\bibnamefont
  {Radu}}, \bibinfo {author} {\bibfnamefont {K.}~\bibnamefont {Vahaplar}},
  \bibinfo {author} {\bibfnamefont {C.}~\bibnamefont {Stamm}}, \bibinfo
  {author} {\bibfnamefont {T.}~\bibnamefont {Kachel}}, \bibinfo {author}
  {\bibfnamefont {N.}~\bibnamefont {Pontius}}, \bibinfo {author} {\bibfnamefont
  {H.~A.}\ \bibnamefont {D{\"u}rr}}, \bibinfo {author} {\bibfnamefont {T.~A.}\
  \bibnamefont {Ostler}}, \bibinfo {author} {\bibfnamefont {J.}~\bibnamefont
  {Barker}}, \bibinfo {author} {\bibfnamefont {R.~F.}\ \bibnamefont {Evans}},
  \bibinfo {author} {\bibfnamefont {R.~W.}\ \bibnamefont {Chantrell}},  \emph
  {et~al.},\ }\href@noop {} {\bibfield  {journal} {\bibinfo  {journal}
  {Nature}\ }\textbf {\bibinfo {volume} {472}},\ \bibinfo {pages} {205}
  (\bibinfo {year} {2011})}\BibitemShut {NoStop}%
\bibitem [{\citenamefont {Mentink}\ \emph {et~al.}(2012)\citenamefont
  {Mentink}, \citenamefont {Hellsvik}, \citenamefont {Afanasiev}, \citenamefont
  {Ivanov}, \citenamefont {Kirilyuk}, \citenamefont {Kimel}, \citenamefont
  {Eriksson}, \citenamefont {Katsnelson},\ and\ \citenamefont
  {Rasing}}]{Mentink.2012}%
  \BibitemOpen
  \bibfield  {author} {\bibinfo {author} {\bibfnamefont {J.~H.}\ \bibnamefont
  {Mentink}}, \bibinfo {author} {\bibfnamefont {J.}~\bibnamefont {Hellsvik}},
  \bibinfo {author} {\bibfnamefont {D.~V.}\ \bibnamefont {Afanasiev}}, \bibinfo
  {author} {\bibfnamefont {B.~A.}\ \bibnamefont {Ivanov}}, \bibinfo {author}
  {\bibfnamefont {A.}~\bibnamefont {Kirilyuk}}, \bibinfo {author}
  {\bibfnamefont {A.~V.}\ \bibnamefont {Kimel}}, \bibinfo {author}
  {\bibfnamefont {O.}~\bibnamefont {Eriksson}}, \bibinfo {author}
  {\bibfnamefont {M.~I.}\ \bibnamefont {Katsnelson}}, \ and\ \bibinfo {author}
  {\bibfnamefont {T.}~\bibnamefont {Rasing}},\ }\href@noop {} {\bibfield
  {journal} {\bibinfo  {journal} {Phys.\ Rev.\ Lett.}\ }\textbf {\bibinfo
  {volume} {108}},\ \bibinfo {pages} {057202} (\bibinfo {year}
  {2012})}\BibitemShut {NoStop}%
\bibitem [{\citenamefont {Ostler}\ \emph {et~al.}(2012)\citenamefont {Ostler},
  \citenamefont {Barker}, \citenamefont {Evans}, \citenamefont {Chantrell},
  \citenamefont {Atxitia}, \citenamefont {Chubykalo-Fesenko}, \citenamefont
  {{El~Moussaoui}}, \citenamefont {{Le~Guyader}}, \citenamefont {Mengotti},
  \citenamefont {Heyderman} \emph {et~al.}}]{Ostler.2012}%
  \BibitemOpen
  \bibfield  {author} {\bibinfo {author} {\bibfnamefont {T.~A.}\ \bibnamefont
  {Ostler}}, \bibinfo {author} {\bibfnamefont {J.}~\bibnamefont {Barker}},
  \bibinfo {author} {\bibfnamefont {R.~F.}\ \bibnamefont {Evans}}, \bibinfo
  {author} {\bibfnamefont {R.~W.}\ \bibnamefont {Chantrell}}, \bibinfo {author}
  {\bibfnamefont {U.}~\bibnamefont {Atxitia}}, \bibinfo {author} {\bibfnamefont
  {O.}~\bibnamefont {Chubykalo-Fesenko}}, \bibinfo {author} {\bibfnamefont
  {S.}~\bibnamefont {{El~Moussaoui}}}, \bibinfo {author} {\bibfnamefont
  {L.}~\bibnamefont {{Le~Guyader}}}, \bibinfo {author} {\bibfnamefont
  {E.}~\bibnamefont {Mengotti}}, \bibinfo {author} {\bibfnamefont {L.~J.}\
  \bibnamefont {Heyderman}},  \emph {et~al.},\ }\href@noop {} {\bibfield
  {journal} {\bibinfo  {journal} {Nat.\ Commun.}\ }\textbf {\bibinfo {volume}
  {3}},\ \bibinfo {pages} {666} (\bibinfo {year} {2012})}\BibitemShut {NoStop}%
\bibitem [{\citenamefont {Wienholdt}\ \emph {et~al.}(2013)\citenamefont
  {Wienholdt}, \citenamefont {Hinzke}, \citenamefont {Carva}, \citenamefont
  {Oppeneer},\ and\ \citenamefont {Nowak}}]{Wienholdt.2013}%
  \BibitemOpen
  \bibfield  {author} {\bibinfo {author} {\bibfnamefont {S.}~\bibnamefont
  {Wienholdt}}, \bibinfo {author} {\bibfnamefont {D.}~\bibnamefont {Hinzke}},
  \bibinfo {author} {\bibfnamefont {K.}~\bibnamefont {Carva}}, \bibinfo
  {author} {\bibfnamefont {P.~M.}\ \bibnamefont {Oppeneer}}, \ and\ \bibinfo
  {author} {\bibfnamefont {U.}~\bibnamefont {Nowak}},\ }\href@noop {}
  {\bibfield  {journal} {\bibinfo  {journal} {Physical Review B}\ }\textbf
  {\bibinfo {volume} {88}},\ \bibinfo {pages} {020406} (\bibinfo {year}
  {2013})}\BibitemShut {NoStop}%
\bibitem [{\citenamefont {Baral}\ and\ \citenamefont
  {Schneider}(2015)}]{Baral:2015fu}%
  \BibitemOpen
  \bibfield  {author} {\bibinfo {author} {\bibfnamefont {A.}~\bibnamefont
  {Baral}}\ and\ \bibinfo {author} {\bibfnamefont {H.~C.}\ \bibnamefont
  {Schneider}},\ }\href@noop {} {\bibfield  {journal} {\bibinfo  {journal}
  {Phys.\ Rev.~B}\ }\textbf {\bibinfo {volume} {91}},\ \bibinfo {pages}
  {100402(R)} (\bibinfo {year} {2015})}\BibitemShut {NoStop}%
\bibitem [{\citenamefont {Mangin}\ \emph {et~al.}(2014)\citenamefont {Mangin},
  \citenamefont {Gottwald}, \citenamefont {Lambert}, \citenamefont {Steil},
  \citenamefont {Uhl{\'\i}{\v r}}, \citenamefont {Pang}, \citenamefont {Hehn},
  \citenamefont {Alebrand}, \citenamefont {Cinchetti}, \citenamefont
  {Malinowski}, \citenamefont {Fainman}, \citenamefont {Aeschlimann},\ and\
  \citenamefont {Fullerton}}]{Mangin2014-bx}%
  \BibitemOpen
  \bibfield  {author} {\bibinfo {author} {\bibfnamefont {S.}~\bibnamefont
  {Mangin}}, \bibinfo {author} {\bibfnamefont {M.}~\bibnamefont {Gottwald}},
  \bibinfo {author} {\bibfnamefont {C.-H.}\ \bibnamefont {Lambert}}, \bibinfo
  {author} {\bibfnamefont {D.}~\bibnamefont {Steil}}, \bibinfo {author}
  {\bibfnamefont {V.}~\bibnamefont {Uhl{\'\i}{\v r}}}, \bibinfo {author}
  {\bibfnamefont {L.}~\bibnamefont {Pang}}, \bibinfo {author} {\bibfnamefont
  {M.}~\bibnamefont {Hehn}}, \bibinfo {author} {\bibfnamefont {S.}~\bibnamefont
  {Alebrand}}, \bibinfo {author} {\bibfnamefont {M.}~\bibnamefont {Cinchetti}},
  \bibinfo {author} {\bibfnamefont {G.}~\bibnamefont {Malinowski}}, \bibinfo
  {author} {\bibfnamefont {Y.}~\bibnamefont {Fainman}}, \bibinfo {author}
  {\bibfnamefont {M.}~\bibnamefont {Aeschlimann}}, \ and\ \bibinfo {author}
  {\bibfnamefont {E.~E.}\ \bibnamefont {Fullerton}},\ }\href@noop {} {\bibfield
   {journal} {\bibinfo  {journal} {Nat. Mater.}\ }\textbf {\bibinfo {volume}
  {13}},\ \bibinfo {pages} {286} (\bibinfo {year} {2014})}\BibitemShut
  {NoStop}%
\bibitem [{\citenamefont {John}\ \emph {et~al.}(2017)\citenamefont {John},
  \citenamefont {Berritta}, \citenamefont {Hinzke}, \citenamefont {M{\"u}ller},
  \citenamefont {Santos}, \citenamefont {Ulrichs}, \citenamefont {Nieves},
  \citenamefont {Walowski}, \citenamefont {Mondal}, \citenamefont
  {Chubykalo-Fesenko}, \citenamefont {McCord}, \citenamefont {Oppeneer},
  \citenamefont {Nowak},\ and\ \citenamefont {M{\"u}nzenberg}}]{John2017-mk}%
  \BibitemOpen
  \bibfield  {author} {\bibinfo {author} {\bibfnamefont {R.}~\bibnamefont
  {John}}, \bibinfo {author} {\bibfnamefont {M.}~\bibnamefont {Berritta}},
  \bibinfo {author} {\bibfnamefont {D.}~\bibnamefont {Hinzke}}, \bibinfo
  {author} {\bibfnamefont {C.}~\bibnamefont {M{\"u}ller}}, \bibinfo {author}
  {\bibfnamefont {T.}~\bibnamefont {Santos}}, \bibinfo {author} {\bibfnamefont
  {H.}~\bibnamefont {Ulrichs}}, \bibinfo {author} {\bibfnamefont
  {P.}~\bibnamefont {Nieves}}, \bibinfo {author} {\bibfnamefont
  {J.}~\bibnamefont {Walowski}}, \bibinfo {author} {\bibfnamefont
  {R.}~\bibnamefont {Mondal}}, \bibinfo {author} {\bibfnamefont
  {O.}~\bibnamefont {Chubykalo-Fesenko}}, \bibinfo {author} {\bibfnamefont
  {J.}~\bibnamefont {McCord}}, \bibinfo {author} {\bibfnamefont {P.~M.}\
  \bibnamefont {Oppeneer}}, \bibinfo {author} {\bibfnamefont {U.}~\bibnamefont
  {Nowak}}, \ and\ \bibinfo {author} {\bibfnamefont {M.}~\bibnamefont
  {M{\"u}nzenberg}},\ }\href@noop {} {\bibfield  {journal} {\bibinfo  {journal}
  {Sci. Rep.}\ }\textbf {\bibinfo {volume} {7}},\ \bibinfo {pages} {4114}
  (\bibinfo {year} {2017})}\BibitemShut {NoStop}%
\bibitem [{\citenamefont {Kimel}\ \emph {et~al.}(2005)\citenamefont {Kimel},
  \citenamefont {Kirilyuk}, \citenamefont {Usachev}, \citenamefont {Pisarev},
  \citenamefont {Balbashov},\ and\ \citenamefont {Rasing}}]{Kimel2005-lq}%
  \BibitemOpen
  \bibfield  {author} {\bibinfo {author} {\bibfnamefont {A.~V.}\ \bibnamefont
  {Kimel}}, \bibinfo {author} {\bibfnamefont {A.}~\bibnamefont {Kirilyuk}},
  \bibinfo {author} {\bibfnamefont {P.~A.}\ \bibnamefont {Usachev}}, \bibinfo
  {author} {\bibfnamefont {R.~V.}\ \bibnamefont {Pisarev}}, \bibinfo {author}
  {\bibfnamefont {A.~M.}\ \bibnamefont {Balbashov}}, \ and\ \bibinfo {author}
  {\bibfnamefont {T.}~\bibnamefont {Rasing}},\ }\href@noop {} {\bibfield
  {journal} {\bibinfo  {journal} {Nature}\ }\textbf {\bibinfo {volume} {435}},\
  \bibinfo {pages} {655} (\bibinfo {year} {2005})}\BibitemShut {NoStop}%
\bibitem [{\citenamefont {Kirilyuk}\ \emph {et~al.}(2010)\citenamefont
  {Kirilyuk}, \citenamefont {Kimel},\ and\ \citenamefont
  {Rasing}}]{Kirilyuk2010-wx}%
  \BibitemOpen
  \bibfield  {author} {\bibinfo {author} {\bibfnamefont {A.}~\bibnamefont
  {Kirilyuk}}, \bibinfo {author} {\bibfnamefont {A.~V.}\ \bibnamefont {Kimel}},
  \ and\ \bibinfo {author} {\bibfnamefont {T.}~\bibnamefont {Rasing}},\
  }\href@noop {} {\bibfield  {journal} {\bibinfo  {journal} {Rev. Mod. Phys.}\
  }\textbf {\bibinfo {volume} {82}},\ \bibinfo {pages} {2731} (\bibinfo {year}
  {2010})}\BibitemShut {NoStop}%
\bibitem [{\citenamefont {Alebrand}\ \emph {et~al.}(2012)\citenamefont
  {Alebrand}, \citenamefont {Hassdenteufel}, \citenamefont {Steil},
  \citenamefont {Cinchetti},\ and\ \citenamefont
  {Aeschlimann}}]{Alebrand2012-dj}%
  \BibitemOpen
  \bibfield  {author} {\bibinfo {author} {\bibfnamefont {S.}~\bibnamefont
  {Alebrand}}, \bibinfo {author} {\bibfnamefont {A.}~\bibnamefont
  {Hassdenteufel}}, \bibinfo {author} {\bibfnamefont {D.}~\bibnamefont
  {Steil}}, \bibinfo {author} {\bibfnamefont {M.}~\bibnamefont {Cinchetti}}, \
  and\ \bibinfo {author} {\bibfnamefont {M.}~\bibnamefont {Aeschlimann}},\
  }\href@noop {} {\bibfield  {journal} {\bibinfo  {journal} {Phys. Rev. B}\
  }\textbf {\bibinfo {volume} {85}},\ \bibinfo {pages} {092401} (\bibinfo
  {year} {2012})}\BibitemShut {NoStop}%
\bibitem [{\citenamefont {Hassdenteufel}\ \emph {et~al.}(2015)\citenamefont
  {Hassdenteufel}, \citenamefont {Schmidt}, \citenamefont {Schubert},
  \citenamefont {Hebler}, \citenamefont {Helm}, \citenamefont {Albrecht},\ and\
  \citenamefont {Bratschitsch}}]{Hassdenteufel2015-om}%
  \BibitemOpen
  \bibfield  {author} {\bibinfo {author} {\bibfnamefont {A.}~\bibnamefont
  {Hassdenteufel}}, \bibinfo {author} {\bibfnamefont {J.}~\bibnamefont
  {Schmidt}}, \bibinfo {author} {\bibfnamefont {C.}~\bibnamefont {Schubert}},
  \bibinfo {author} {\bibfnamefont {B.}~\bibnamefont {Hebler}}, \bibinfo
  {author} {\bibfnamefont {M.}~\bibnamefont {Helm}}, \bibinfo {author}
  {\bibfnamefont {M.}~\bibnamefont {Albrecht}}, \ and\ \bibinfo {author}
  {\bibfnamefont {R.}~\bibnamefont {Bratschitsch}},\ }\href@noop {} {\bibfield
  {journal} {\bibinfo  {journal} {Phys. Rev. B Condens. Matter}\ }\textbf
  {\bibinfo {volume} {91}},\ \bibinfo {pages} {104431} (\bibinfo {year}
  {2015})}\BibitemShut {NoStop}%
\bibitem [{\citenamefont {Pitaevskii}(1961)}]{Pitaevskii1961-qd}%
  \BibitemOpen
  \bibfield  {author} {\bibinfo {author} {\bibfnamefont {L.~P.}\ \bibnamefont
  {Pitaevskii}},\ }\href@noop {} {\bibfield  {journal} {\bibinfo  {journal}
  {Sov. Phys. JETP}\ }\textbf {\bibinfo {volume} {12}},\ \bibinfo {pages}
  {1008} (\bibinfo {year} {1961})}\BibitemShut {NoStop}%
\bibitem [{\citenamefont {Pershan}\ \emph {et~al.}(1966)\citenamefont
  {Pershan}, \citenamefont {Van~der Ziel},\ and\ \citenamefont
  {Malmstrom}}]{Pershan1966-wf}%
  \BibitemOpen
  \bibfield  {author} {\bibinfo {author} {\bibfnamefont {P.}~\bibnamefont
  {Pershan}}, \bibinfo {author} {\bibfnamefont {J.}~\bibnamefont {Van~der
  Ziel}}, \ and\ \bibinfo {author} {\bibfnamefont {L.}~\bibnamefont
  {Malmstrom}},\ }\href@noop {} {\bibfield  {journal} {\bibinfo  {journal}
  {Physical Review}\ } (\bibinfo {year} {1966})}\BibitemShut {NoStop}%
\bibitem [{\citenamefont {Hertel}(2006)}]{Hertel.2006}%
  \BibitemOpen
  \bibfield  {author} {\bibinfo {author} {\bibfnamefont {R.}~\bibnamefont
  {Hertel}},\ }\href@noop {} {\bibfield  {journal} {\bibinfo  {journal}
  {Journal of magnetism and magnetic materials}\ }\textbf {\bibinfo {volume}
  {303}},\ \bibinfo {pages} {L1} (\bibinfo {year} {2006})}\BibitemShut
  {NoStop}%
\bibitem [{\citenamefont {Kurkin}\ \emph {et~al.}(2008)\citenamefont {Kurkin},
  \citenamefont {Bakulina},\ and\ \citenamefont {Pisarev}}]{Kurkin2008-js}%
  \BibitemOpen
  \bibfield  {author} {\bibinfo {author} {\bibfnamefont {M.~I.}\ \bibnamefont
  {Kurkin}}, \bibinfo {author} {\bibfnamefont {N.~B.}\ \bibnamefont
  {Bakulina}}, \ and\ \bibinfo {author} {\bibfnamefont {R.~V.}\ \bibnamefont
  {Pisarev}},\ }\href@noop {} {\bibfield  {journal} {\bibinfo  {journal} {Phys.
  Rev. B Condens. Matter}\ }\textbf {\bibinfo {volume} {78}},\ \bibinfo {pages}
  {134430} (\bibinfo {year} {2008})}\BibitemShut {NoStop}%
\bibitem [{\citenamefont {Woodford}(2009)}]{Woodford2009-kf}%
  \BibitemOpen
  \bibfield  {author} {\bibinfo {author} {\bibfnamefont {S.~R.}\ \bibnamefont
  {Woodford}},\ }\href@noop {} {\bibfield  {journal} {\bibinfo  {journal}
  {Phys. Rev. B Condens. Matter}\ }\textbf {\bibinfo {volume} {79}},\ \bibinfo
  {pages} {212412} (\bibinfo {year} {2009})}\BibitemShut {NoStop}%
\bibitem [{\citenamefont {Taguchi}\ and\ \citenamefont
  {Tatara}(2011)}]{Taguchi2011-bi}%
  \BibitemOpen
  \bibfield  {author} {\bibinfo {author} {\bibfnamefont {K.}~\bibnamefont
  {Taguchi}}\ and\ \bibinfo {author} {\bibfnamefont {G.}~\bibnamefont
  {Tatara}},\ }\href@noop {} {\bibfield  {journal} {\bibinfo  {journal} {Phys.
  Rev. B Condens. Matter}\ }\textbf {\bibinfo {volume} {84}},\ \bibinfo {pages}
  {174433} (\bibinfo {year} {2011})}\BibitemShut {NoStop}%
\bibitem [{\citenamefont {Hertel}\ and\ \citenamefont
  {F{\"a}hnle}(2015)}]{Hertel2015-oa}%
  \BibitemOpen
  \bibfield  {author} {\bibinfo {author} {\bibfnamefont {R.}~\bibnamefont
  {Hertel}}\ and\ \bibinfo {author} {\bibfnamefont {M.}~\bibnamefont
  {F{\"a}hnle}},\ }\href@noop {} {\bibfield  {journal} {\bibinfo  {journal}
  {Phys. Rev. B Condens. Matter}\ }\textbf {\bibinfo {volume} {91}},\ \bibinfo
  {pages} {020411} (\bibinfo {year} {2015})}\BibitemShut {NoStop}%
\bibitem [{\citenamefont {Edelstein}(1998)}]{Edelstein1998-lw}%
  \BibitemOpen
  \bibfield  {author} {\bibinfo {author} {\bibfnamefont {V.~M.}\ \bibnamefont
  {Edelstein}},\ }\href@noop {} {\bibfield  {journal} {\bibinfo  {journal}
  {Phys. Rev. Lett.}\ }\textbf {\bibinfo {volume} {80}},\ \bibinfo {pages}
  {5766} (\bibinfo {year} {1998})}\BibitemShut {NoStop}%
\bibitem [{\citenamefont {Qaiumzadeh}\ and\ \citenamefont
  {Titov}(2016)}]{Qaiumzadeh2016-pd}%
  \BibitemOpen
  \bibfield  {author} {\bibinfo {author} {\bibfnamefont {A.}~\bibnamefont
  {Qaiumzadeh}}\ and\ \bibinfo {author} {\bibfnamefont {M.}~\bibnamefont
  {Titov}},\ }\href@noop {} {\bibfield  {journal} {\bibinfo  {journal} {Phys.
  Rev. B}\ }\textbf {\bibinfo {volume} {94}},\ \bibinfo {pages} {014425}
  (\bibinfo {year} {2016})}\BibitemShut {NoStop}%
\bibitem [{\citenamefont {Zheludev}\ \emph {et~al.}(1994)\citenamefont
  {Zheludev}, \citenamefont {Brummell}, \citenamefont {Harley}, \citenamefont
  {Malinowski}, \citenamefont {Popov}, \citenamefont {Ashenford},\ and\
  \citenamefont {Lunn}}]{Zheludev1994-im}%
  \BibitemOpen
  \bibfield  {author} {\bibinfo {author} {\bibfnamefont {N.~I.}\ \bibnamefont
  {Zheludev}}, \bibinfo {author} {\bibfnamefont {M.~A.}\ \bibnamefont
  {Brummell}}, \bibinfo {author} {\bibfnamefont {R.~T.}\ \bibnamefont
  {Harley}}, \bibinfo {author} {\bibfnamefont {A.}~\bibnamefont {Malinowski}},
  \bibinfo {author} {\bibfnamefont {S.~V.}\ \bibnamefont {Popov}}, \bibinfo
  {author} {\bibfnamefont {D.~E.}\ \bibnamefont {Ashenford}}, \ and\ \bibinfo
  {author} {\bibfnamefont {B.}~\bibnamefont {Lunn}},\ }\href@noop {} {\bibfield
   {journal} {\bibinfo  {journal} {Solid State Commun.}\ }\textbf {\bibinfo
  {volume} {89}},\ \bibinfo {pages} {823} (\bibinfo {year} {1994})}\BibitemShut
  {NoStop}%
\bibitem [{\citenamefont {Mikhaylovskiy}\ \emph {et~al.}(2012)\citenamefont
  {Mikhaylovskiy}, \citenamefont {Hendry},\ and\ \citenamefont
  {Kruglyak}}]{Mikhaylovskiy2012-vs}%
  \BibitemOpen
  \bibfield  {author} {\bibinfo {author} {\bibfnamefont {R.~V.}\ \bibnamefont
  {Mikhaylovskiy}}, \bibinfo {author} {\bibfnamefont {E.}~\bibnamefont
  {Hendry}}, \ and\ \bibinfo {author} {\bibfnamefont {V.~V.}\ \bibnamefont
  {Kruglyak}},\ }\href@noop {} {\bibfield  {journal} {\bibinfo  {journal}
  {Phys. Rev. B Condens. Matter}\ }\textbf {\bibinfo {volume} {86}},\ \bibinfo
  {pages} {100405} (\bibinfo {year} {2012})}\BibitemShut {NoStop}%
\bibitem [{\citenamefont {Battiato}\ \emph {et~al.}(2014)\citenamefont
  {Battiato}, \citenamefont {Barbalinardo},\ and\ \citenamefont
  {Oppeneer}}]{Battiato2014-mu}%
  \BibitemOpen
  \bibfield  {author} {\bibinfo {author} {\bibfnamefont {M.}~\bibnamefont
  {Battiato}}, \bibinfo {author} {\bibfnamefont {G.}~\bibnamefont
  {Barbalinardo}}, \ and\ \bibinfo {author} {\bibfnamefont {P.~M.}\
  \bibnamefont {Oppeneer}},\ }\href@noop {} {\bibfield  {journal} {\bibinfo
  {journal} {Phys. Rev. B Condens. Matter}\ }\textbf {\bibinfo {volume} {89}},\
  \bibinfo {pages} {014413} (\bibinfo {year} {2014})}\BibitemShut {NoStop}%
\bibitem [{\citenamefont {Berritta}\ \emph
  {et~al.}(2016{\natexlab{a}})\citenamefont {Berritta}, \citenamefont {Mondal},
  \citenamefont {Carva},\ and\ \citenamefont {Oppeneer}}]{Berritta2016-lz}%
  \BibitemOpen
  \bibfield  {author} {\bibinfo {author} {\bibfnamefont {M.}~\bibnamefont
  {Berritta}}, \bibinfo {author} {\bibfnamefont {R.}~\bibnamefont {Mondal}},
  \bibinfo {author} {\bibfnamefont {K.}~\bibnamefont {Carva}}, \ and\ \bibinfo
  {author} {\bibfnamefont {P.~M.}\ \bibnamefont {Oppeneer}},\ }\href@noop {}
  {\bibfield  {journal} {\bibinfo  {journal} {Phys. Rev. Lett.}\ }\textbf
  {\bibinfo {volume} {117}},\ \bibinfo {pages} {137203} (\bibinfo {year}
  {2016}{\natexlab{a}})}\BibitemShut {NoStop}%
\bibitem [{\citenamefont {Popova}\ \emph {et~al.}(2012)\citenamefont {Popova},
  \citenamefont {Bringer},\ and\ \citenamefont {Bl{\"u}gel}}]{Popova2012-zl}%
  \BibitemOpen
  \bibfield  {author} {\bibinfo {author} {\bibfnamefont {D.}~\bibnamefont
  {Popova}}, \bibinfo {author} {\bibfnamefont {A.}~\bibnamefont {Bringer}}, \
  and\ \bibinfo {author} {\bibfnamefont {S.}~\bibnamefont {Bl{\"u}gel}},\
  }\href@noop {} {\bibfield  {journal} {\bibinfo  {journal} {Phys. Rev. B}\
  }\textbf {\bibinfo {volume} {85}},\ \bibinfo {pages} {094419} (\bibinfo
  {year} {2012})}\BibitemShut {NoStop}%
\bibitem [{\citenamefont {Qaiumzadeh}\ \emph {et~al.}(2013)\citenamefont
  {Qaiumzadeh}, \citenamefont {Bauer},\ and\ \citenamefont
  {Brataas}}]{Qaiumzadeh2013-cy}%
  \BibitemOpen
  \bibfield  {author} {\bibinfo {author} {\bibfnamefont {A.}~\bibnamefont
  {Qaiumzadeh}}, \bibinfo {author} {\bibfnamefont {G.~E.~W.}\ \bibnamefont
  {Bauer}}, \ and\ \bibinfo {author} {\bibfnamefont {A.}~\bibnamefont
  {Brataas}},\ }\href@noop {} {\bibfield  {journal} {\bibinfo  {journal} {Phys.
  Rev. B Condens. Matter}\ }\textbf {\bibinfo {volume} {88}},\ \bibinfo {pages}
  {064416} (\bibinfo {year} {2013})}\BibitemShut {NoStop}%
\bibitem [{\citenamefont {Haug}\ and\ \citenamefont
  {Koch}(2009)}]{haug2009quantum}%
  \BibitemOpen
  \bibfield  {author} {\bibinfo {author} {\bibfnamefont {H.}~\bibnamefont
  {Haug}}\ and\ \bibinfo {author} {\bibfnamefont {S.~W.}\ \bibnamefont
  {Koch}},\ }\href@noop {} {\emph {\bibinfo {title} {Quantum theory of the
  optical and electronic properties of semiconductors}}}\ (\bibinfo
  {publisher} {World Scientific Publishing Co Inc},\ \bibinfo {year}
  {2009})\BibitemShut {NoStop}%
\bibitem [{\citenamefont {Scholl}\ \emph {et~al.}(2018)\citenamefont {Scholl},
  \citenamefont {Vollmar},\ and\ \citenamefont
  {Schneider}}]{Scholl-unpublished}%
  \BibitemOpen
  \bibfield  {author} {\bibinfo {author} {\bibfnamefont {C.}~\bibnamefont
  {Scholl}}, \bibinfo {author} {\bibfnamefont {S.}~\bibnamefont {Vollmar}}, \
  and\ \bibinfo {author} {\bibfnamefont {H.~C.}\ \bibnamefont {Schneider}},\
  }\href@noop {} {\enquote {\bibinfo {title} {Light-induced magnetization
  dynamics in a ferroamgnetic rashba model},}\ } (\bibinfo {year} {2018}),\
  \bibinfo {note} {unpublished}\BibitemShut {NoStop}%
\bibitem [{\citenamefont {Popova}\ \emph {et~al.}(2011)\citenamefont {Popova},
  \citenamefont {Bringer},\ and\ \citenamefont {Bl{\"u}gel}}]{Popova2011-jr}%
  \BibitemOpen
  \bibfield  {author} {\bibinfo {author} {\bibfnamefont {D.}~\bibnamefont
  {Popova}}, \bibinfo {author} {\bibfnamefont {A.}~\bibnamefont {Bringer}}, \
  and\ \bibinfo {author} {\bibfnamefont {S.}~\bibnamefont {Bl{\"u}gel}},\
  }\href@noop {} {\bibfield  {journal} {\bibinfo  {journal} {Phys. Rev. B
  Condens. Matter}\ }\textbf {\bibinfo {volume} {84}},\ \bibinfo {pages}
  {214421} (\bibinfo {year} {2011})}\BibitemShut {NoStop}%
\bibitem [{\citenamefont {Sussman}(2011)}]{Sussman.2011}%
  \BibitemOpen
  \bibfield  {author} {\bibinfo {author} {\bibfnamefont {B.~J.}\ \bibnamefont
  {Sussman}},\ }\href {\doibase 10.1119/1.3553018} {\bibfield  {journal}
  {\bibinfo  {journal} {American Journal of Physics}\ }\textbf {\bibinfo
  {volume} {79}},\ \bibinfo {pages} {477} (\bibinfo {year} {2011})},\ \Eprint
  {http://arxiv.org/abs/https://doi.org/10.1119/1.3553018}
  {https://doi.org/10.1119/1.3553018} \BibitemShut {NoStop}%
\bibitem [{\citenamefont {Krau{\ss}}\ \emph {et~al.}(2009)\citenamefont
  {Krau{\ss}}, \citenamefont {Roth}, \citenamefont {Alebrand}, \citenamefont
  {Steil}, \citenamefont {Cinchetti}, \citenamefont {Aeschlimann},\ and\
  \citenamefont {Schneider}}]{Krauss.2009}%
  \BibitemOpen
  \bibfield  {author} {\bibinfo {author} {\bibfnamefont {M.}~\bibnamefont
  {Krau{\ss}}}, \bibinfo {author} {\bibfnamefont {T.}~\bibnamefont {Roth}},
  \bibinfo {author} {\bibfnamefont {S.}~\bibnamefont {Alebrand}}, \bibinfo
  {author} {\bibfnamefont {D.}~\bibnamefont {Steil}}, \bibinfo {author}
  {\bibfnamefont {M.}~\bibnamefont {Cinchetti}}, \bibinfo {author}
  {\bibfnamefont {M.}~\bibnamefont {Aeschlimann}}, \ and\ \bibinfo {author}
  {\bibfnamefont {H.~C.}\ \bibnamefont {Schneider}},\ }\href@noop {} {\bibfield
   {journal} {\bibinfo  {journal} {Phys. Rev. B}\ }\textbf {\bibinfo {volume}
  {80}},\ \bibinfo {pages} {180407} (\bibinfo {year} {2009})}\BibitemShut
  {NoStop}%
\bibitem [{\citenamefont {M{\"{u}}ller}\ \emph {et~al.}(2013)\citenamefont
  {M{\"{u}}ller}, \citenamefont {Baral}, \citenamefont {Vollmar}, \citenamefont
  {Cinchetti}, \citenamefont {Aeschlimann}, \citenamefont {Schneider},\ and\
  \citenamefont {Rethfeld}}]{Mueller.2013}%
  \BibitemOpen
  \bibfield  {author} {\bibinfo {author} {\bibfnamefont {B.~Y.}\ \bibnamefont
  {M{\"{u}}ller}}, \bibinfo {author} {\bibfnamefont {A.}~\bibnamefont {Baral}},
  \bibinfo {author} {\bibfnamefont {S.}~\bibnamefont {Vollmar}}, \bibinfo
  {author} {\bibfnamefont {M.}~\bibnamefont {Cinchetti}}, \bibinfo {author}
  {\bibfnamefont {M.}~\bibnamefont {Aeschlimann}}, \bibinfo {author}
  {\bibfnamefont {H.~C.}\ \bibnamefont {Schneider}}, \ and\ \bibinfo {author}
  {\bibfnamefont {B.}~\bibnamefont {Rethfeld}},\ }\href@noop {} {\bibfield
  {journal} {\bibinfo  {journal} {Phys.\ Rev.~B}\ }\textbf {\bibinfo {volume}
  {111}},\ \bibinfo {pages} {167204} (\bibinfo {year} {2013})}\BibitemShut
  {NoStop}%
\bibitem [{\citenamefont {Leckron}\ \emph {et~al.}(2017)\citenamefont
  {Leckron}, \citenamefont {Vollmar},\ and\ \citenamefont
  {Schneider}}]{Leckron.2017}%
  \BibitemOpen
  \bibfield  {author} {\bibinfo {author} {\bibfnamefont {K.}~\bibnamefont
  {Leckron}}, \bibinfo {author} {\bibfnamefont {S.}~\bibnamefont {Vollmar}}, \
  and\ \bibinfo {author} {\bibfnamefont {H.~C.}\ \bibnamefont {Schneider}},\
  }\href {\doibase 10.1103/PhysRevB.96.140408} {\bibfield  {journal} {\bibinfo
  {journal} {Phys. Rev. B}\ }\textbf {\bibinfo {volume} {96}},\ \bibinfo
  {pages} {140408} (\bibinfo {year} {2017})}\BibitemShut {NoStop}%
\bibitem [{\citenamefont {El~Hadri}\ \emph {et~al.}(2016)\citenamefont
  {El~Hadri}, \citenamefont {Pirro}, \citenamefont {Lambert}, \citenamefont
  {Petit-Watelot}, \citenamefont {Quessab}, \citenamefont {Hehn}, \citenamefont
  {Montaigne}, \citenamefont {Malinowski},\ and\ \citenamefont
  {Mangin}}]{El_Hadri2016-xa}%
  \BibitemOpen
  \bibfield  {author} {\bibinfo {author} {\bibfnamefont {M.~S.}\ \bibnamefont
  {El~Hadri}}, \bibinfo {author} {\bibfnamefont {P.}~\bibnamefont {Pirro}},
  \bibinfo {author} {\bibfnamefont {C.~H.}\ \bibnamefont {Lambert}}, \bibinfo
  {author} {\bibfnamefont {S.}~\bibnamefont {Petit-Watelot}}, \bibinfo {author}
  {\bibfnamefont {Y.}~\bibnamefont {Quessab}}, \bibinfo {author} {\bibfnamefont
  {M.}~\bibnamefont {Hehn}}, \bibinfo {author} {\bibfnamefont {F.}~\bibnamefont
  {Montaigne}}, \bibinfo {author} {\bibfnamefont {G.}~\bibnamefont
  {Malinowski}}, \ and\ \bibinfo {author} {\bibfnamefont {S.}~\bibnamefont
  {Mangin}},\ }\href@noop {} {\bibfield  {journal} {\bibinfo  {journal} {Phys.
  Rev. B: Condens. Matter Mater. Phys.}\ }\textbf {\bibinfo {volume} {94}},\
  \bibinfo {pages} {064412} (\bibinfo {year} {2016})}\BibitemShut {NoStop}%
\bibitem [{\citenamefont {Berritta}\ \emph
  {et~al.}(2016{\natexlab{b}})\citenamefont {Berritta}, \citenamefont {Mondal},
  \citenamefont {Carva},\ and\ \citenamefont {Oppeneer}}]{Berritta.2016}%
  \BibitemOpen
  \bibfield  {author} {\bibinfo {author} {\bibfnamefont {M.}~\bibnamefont
  {Berritta}}, \bibinfo {author} {\bibfnamefont {R.}~\bibnamefont {Mondal}},
  \bibinfo {author} {\bibfnamefont {K.}~\bibnamefont {Carva}}, \ and\ \bibinfo
  {author} {\bibfnamefont {P.~M.}\ \bibnamefont {Oppeneer}},\ }\href {\doibase
  10.1103/PhysRevLett.117.137203} {\bibfield  {journal} {\bibinfo  {journal}
  {Phys. Rev. Lett.}\ }\textbf {\bibinfo {volume} {117}},\ \bibinfo {pages}
  {137203} (\bibinfo {year} {2016}{\natexlab{b}})}\BibitemShut {NoStop}%
\bibitem [{\citenamefont {Agranat}\ \emph {et~al.}(1986)\citenamefont
  {Agranat}, \citenamefont {Ashikov}, \citenamefont {Granovskii},\ and\
  \citenamefont {Rukman}}]{Agranat.1986}%
  \BibitemOpen
  \bibfield  {author} {\bibinfo {author} {\bibfnamefont {M.~B.}\ \bibnamefont
  {Agranat}}, \bibinfo {author} {\bibfnamefont {S.~I.}\ \bibnamefont
  {Ashikov}}, \bibinfo {author} {\bibfnamefont {A.~B.}\ \bibnamefont
  {Granovskii}}, \ and\ \bibinfo {author} {\bibfnamefont {G.~I.}\ \bibnamefont
  {Rukman}},\ }\href@noop {} {\bibfield  {journal} {\bibinfo  {journal} {Zh.
  Eksp. Teor. Fiz}\ }\textbf {\bibinfo {volume} {86}},\ \bibinfo {pages} {1376}
  (\bibinfo {year} {1986})}\BibitemShut {NoStop}%
\end{thebibliography}

\end{document}